\documentclass[11pt,twoside]{article}

\usepackage{asp2004}
\usepackage{epsf}
\usepackage{psfig}
\usepackage{lscape}

\begin{document}
\title{MUSYC: A Deep Square Degree Survey of the Formation and Evolution of Galaxies and Supermassive Black Holes  }   %
\author{Eric Gawiser and the MUSYC Collaboration}   %
\affil{NSF Astronomy \& Astrophysics Postdoctoral Fellow (AAPF), Yale 
Astronomy Department and Yale Center for Astronomy \& Astrophysics, 
PO Box 208101, New Haven, CT 06520-8101}    %

\begin{abstract} %
The Multiwavelength Survey by Yale-Chile (MUSYC) is optimized for the 
study of galaxies at $z=3$, AGN demographics, and Galactic structure.  
MUSYC 
consists of deep optical (UBVRIz$'$) and near-infrared (JHK) imaging of 
four fields on the sky covering 1.2 square degrees
 to AB limiting depths of 
U,B,V,R=26 and K=22.  Our optical catalog contains 277,341 objects 
detected in BVR images with median seeing of 0.9$''$.  Satellite coverage 
of our fields includes Chandra, XMM, GALEX, HST-ACS, and Spitzer, with the 
Extended Chandra Deep Field South imaged at all of these wavelengths plus 
the radio, making it the premier multiwavelength field on the sky.  Detailed 
follow-up spectroscopy is being performed with VLT+VIMOS, Magellan+IMACS and 
Gemini+GNIRS.  MUSYC provides ideal supporting data for surveys with 
Z-Machines and rich target lists for spectroscopy with ALMA.  We are 
conducting a census of protogalaxies at redshift three (Lyman break 
galaxies, Lyman-alpha emitters, Distant red galaxies, Sub-millimeter 
galaxies and AGN) in order to separate physical properties from 
selection effects.  We discuss measurements of the dark matter halo 
masses and halo occupation numbers of these populations 
and of the total cosmic star formation 
rate at z=3.  
MUSYC publications and data releases are available at 
http://www.astro.yale.edu/MUSYC.
\end{abstract}
\keywords{galaxies:formation,galaxies:high-redshift}

\section{Survey Design}
\label{sec:survey}

The Multiwavelength Survey by Yale-Chile 
is unique for its 
combination of depth and total area, 
for providing the $UBVRIz'JHK$ photometry 
needed for high-quality photometric redshifts over 1.2 square degrees 
of sky, and 
for having additional coverage at X-ray, UV, mid- and far-infrared
wavelengths.  
The primary goal is to study the properties 
and interrelations of galaxies at a single epoch corresponding to 
redshift $\sim3$, using a range of selection techniques.

  Lyman break galaxies at $z\simeq 3$ 
are selected  
through their dropout in $U$-band images 
combined with blue continuua typical of recent star formation at 
$\lambda > 1216${\AA} in the rest-frame \citep{steideletal96, steideletal99, 
steideletal03}
detected in $BVRIz'$.
Imaging depths 
of $U$,$B$,$V$,$R \simeq 26$ were chosen 
to detect the LBGs, whose luminosity 
function has a characteristic magnitude of $m_* = 24.5$ in $R_{\mathrm AB}$, 
and to find their Lyman break decrement in 
the $U$ filter via colors $(U-V)_{\mathrm AB} > 1.2$.    

Lyman $\alpha$ emitters 
at $z\simeq 3$ are selected through additional deep narrow-band imaging using 
a 50{\AA} fwhm filter centered at 5000{\AA}.  These objects can be detected 
in narrow-band imaging and spectroscopy 
by their emission lines, 
allowing us to 
probe to much dimmer continuum magnitudes than possible for Lyman 
break galaxies.  

It has recently become clear that optical selection methods do not 
provide a full census of the galaxy population at $z\sim3$, as they 
miss objects which are faint in the rest-frame ultraviolet 
\citep{franxetal03,daddietal04a}.  With this in mind, MUSYC has 
a comprehensive near-infrared imaging campaign.  The NIR 
imaging comprises two components:
a wide survey
covering the full 
square degree 
and a deep survey
of the central $10'\times 10'$ of each field. 
The $5\sigma$ point source sensitivities of the wide and deep
components are $K_{s,AB}=22.0$ and $K_{s,AB}=23.3$ respectively. 
NIR imaging over the full survey area provides a critical complement 
to optical imaging for breaking degeneracies in photometric redshifts 
and modeling star formation histories.  Deeper $JHK_s$ imaging 
over $10'\times 10'$ subfields opens up an additional window 
into the $z\simeq3$ universe 
as the $J-K$ 
selection technique \citep{franxetal03, vandokkumetal03, vandokkumetal04, 
vandokkumetal06} 
are used to 
find evolved optically-red galaxies at $2<z<4$ through their rest-frame 
Balmer/4000{\AA} break.

In addition to the optical and near-infrared, imaging campaigns 
at other wavelengths and follow-up spectroscopy are integral parts 
of MUSYC.  
X-ray selection is used to study AGN demographics
over the full range of accessible redshifts, $0<z<6$, 
\citep[see][]{liraetal04} 
with Spitzer imaging 
used to detect optically- and X-ray-obscured AGN
\citep{treisteretal04b,lacyetal04}.  
This also allows a census of accreting black holes at $z\simeq 3$ 
in the same fields to study correlations between black hole accretion 
and galaxy properties at this epoch.  

Multiple epochs of optical 
imaging  are being used to conduct a proper motion survey to 
find white dwarfs and brown dwarfs in order to study Galactic structure 
and the local Initial Mass Function
\citep[see][]{altmannetal05}.

The four 
survey fields (see Table~\ref{tab:fields}) were  
chosen to have 
extremely low reddening, H~I column density \citep{bursteinh78}, 
and 100$\mu$m dust emission \citep*{schlegelfd98}.  
Additionally, each field satisfies 
all of the following criteria:
minimal bright foreground sources in the optical and radio, 
high Galactic latitude ($|b|>30$) to reduce stellar density, 
and accessibility from observatories located in Chile.  
The survey fields will be a natural
choice for future observations with ALMA.

\begin{table}[!ht]
\caption{MUSYC Survey Fields and their Galactic extinction, 100$\mu$m emission 
and neutral hydrogen column densities.}
\label{tab:fields}
\smallskip
\begin{center}
{\small 
\begin{tabular}{lccccc}
\tableline
\noalign{\smallskip}
Field&	RA &	DEC &	E(B-V)&	100 $\mu$m Emission  &	N$_H$ \\
     & [J2000]   &   [J2000]& & [MJy/Sr] & [10$^{20}$ cm$^{-2}$]\\
\noalign{\smallskip}
\tableline
ECDF-S&	        03:32:29.0&	-27:48:47&	0.01&	0.40&	0.9\\
SDSS1030+05&	10:30:27.1&	+05:24:55&	0.02&	1.01&	2.3\\
CW1255+01&	12:55:40.0&	+01:07:00&	0.02&	0.81&	1.6\\
EHDF-S&	        22:32:35.6&	-60:47:12&	0.03&	1.37&	1.6\\
\noalign{\smallskip}
\tableline
\end{tabular}
}
\end{center}
\end{table}

\begin{table}[!ht]
\caption{MUSYC optically-selected catalogs and imaging depths (5$\sigma$ point 
source detection in AB magnitudes).  
Depths in $JHK$ are for deep central 
$10'\times10'$ regions in each field, with the surrounding areas covered 
to about one magnitude shallower.}
\label{tab:depths}
\smallskip
\begin{center}
{\small 
\begin{tabular}{lccccccccccc}
\tableline
\noalign{\smallskip}
Field&	 	U&	B&	V&	R&	I&	z$'$&	J&	H&	K&	NB5000\\
\noalign{\smallskip}
\tableline
ECDF-S&	26.0&	26.9&	26.4&	26.4&	24.6&	23.6&	24.3&	23.8&	23.4&	25.5\\
SDSS1030+05&	25.8&	26.0&	26.2&	26.0&	25.4&	23.7&	24.1&	23.9&	23.3&	24.8\\
CW1255+01&	26.0&	26.2&	26.1&	26.0&	25.0&	24.1&	24.0&	22.8&	23.0&	24.4 \\	
EHDF-S&	26.0&	26.1&	26.0&	25.8&	24.7&	23.6&	24.3&	23.4&	23.4&	24.1	\\
\noalign{\smallskip}
\tableline
\end{tabular}
}
\end{center}
\end{table}

\section{Results}
\label{sec:results}

The data reduction and photometry methods used to generate 
optically-selected catalogs are described by \citet{gawiseretal06a}.  
The corrected aperture (APCORR) fluxes yield the optimal point source 
detection depths listed in Table \ref{tab:depths} for the 
277,341 objects in our optical catalog covering 1.2 square degrees.  

\begin{figure}[!ht]
\plottwo{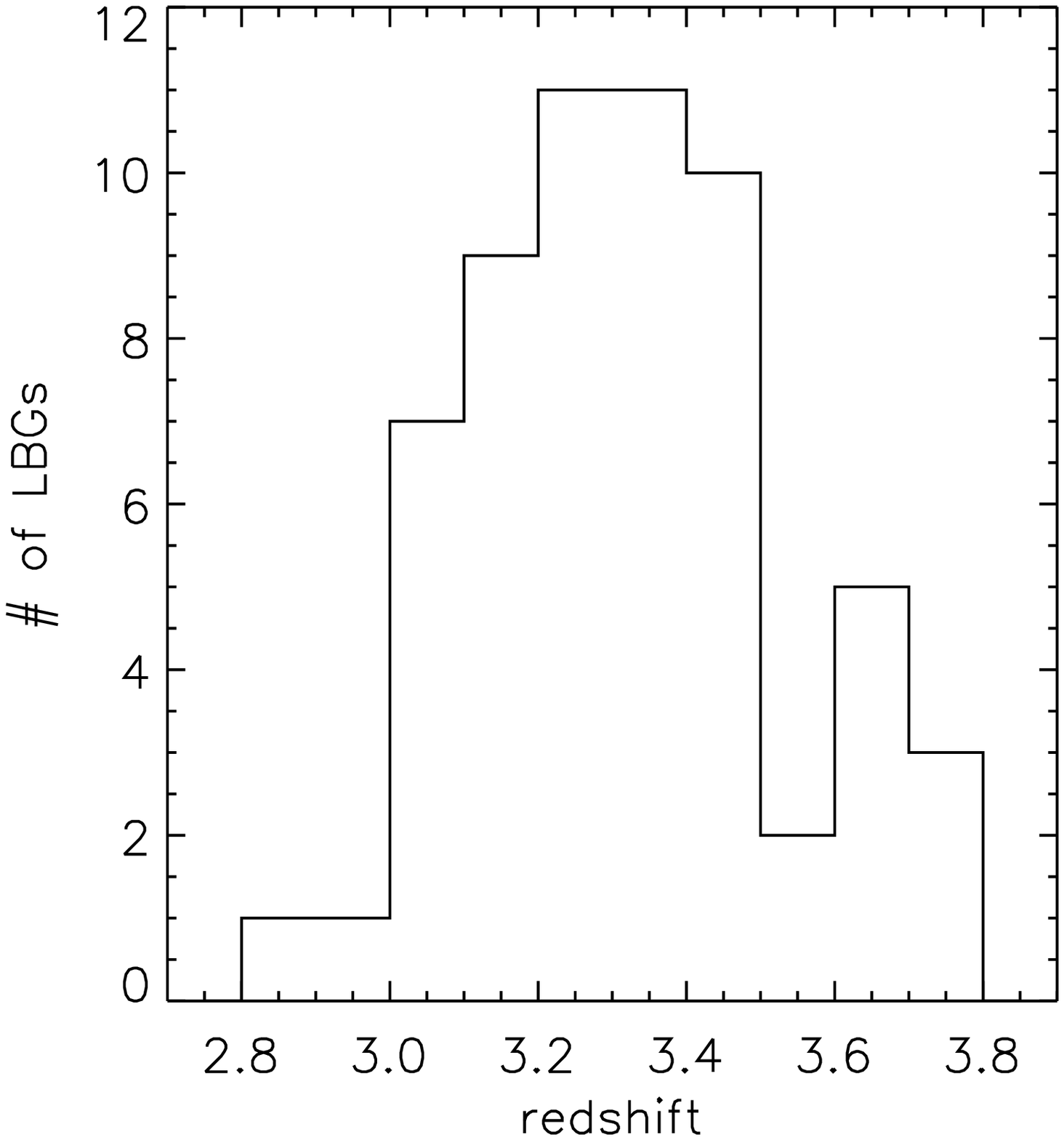}{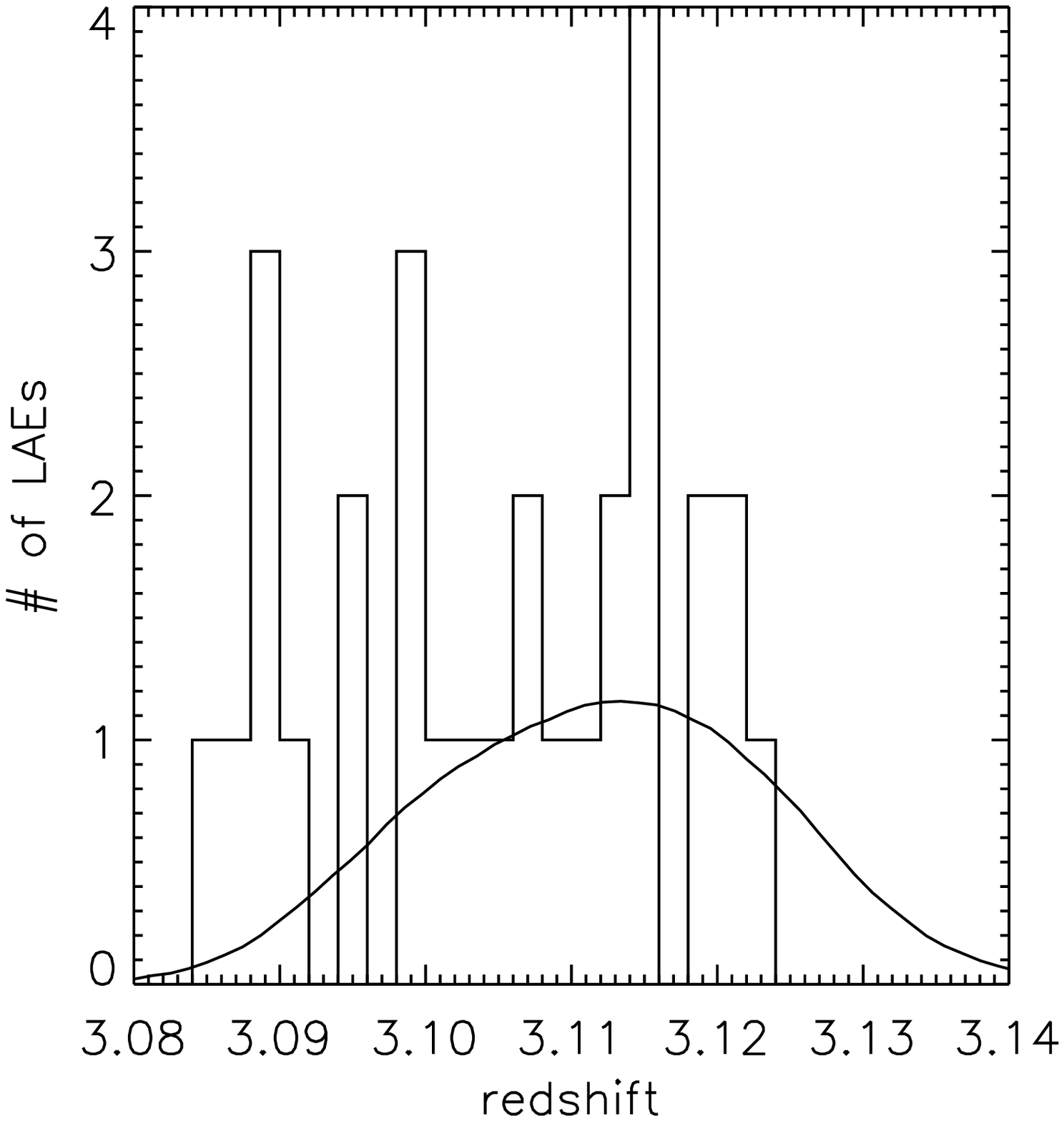}
\caption{Histograms of redshifts for 
60 Lyman break galaxies (left panel) and 
29 Lyman Alpha Emitting galaxies (right panel).
The solid curve in the right 
panel shows the expected redshift distribution calculated by convolving the 
filter 
response with the observed equivalent width 
distribution.} 
\label{fig:hist} 
\end{figure}

Our color selection and clustering analysis of Lyman break galaxies 
(LBGs) is described by \citet{gawiseretal06a}, and our color selection and 
stellar population analysis of Lyman alpha emitting galaxies (LAEs) 
is described by \citet{gawiseretal06b}.  Figure \ref{fig:hist} shows the 
redshift distributions measured  from MUSYC spectroscopy with 
Magellan+IMACS.

The total cosmic star formation rate density 
supplied by spectroscopically confirmed populations 
at $z=3$ 
is 0.2 M$_\odot$ 
yr$^{-1}$  Mpc$^{-3}$.    
Lyman break galaxies only supply half of this, 
with most of the rest supplied by 
Distant Red Galaxies \citep{webbetal06}, 
Sub-millimeter Galaxies \citep{chapmanetal05} 
and Damped Lyman $\alpha$ Absorbers 
(\citealp*{wolfegp03}; for a review, see \citealp*{wolfegp05}).

Preliminary results of our clustering analysis imply that LBGs 
have a halo occupation number close to 1 i.e. each available dark matter 
halo with mass above $3\times10^{11}$M$_\odot$ contains one LBG.  
However, AGN and LAEs have 
halo occupation numbers around 0.1 i.e. only one in ten dark 
matter halos in the mass range hosting these objects are actually found to 
contain one of them.  This implies that 
the duty cycles for active 
accretion onto supermassive black holes and for dust-free star formation
in these objects  
are roughly 10\%.  
All measured protogalaxy families are found to 
have dark matter halo masses of at least 
10$^{11}$M$_\odot$ \citep{gawiser05}, 
implying that 
galaxy formation may have been suppressed sufficiently in typical 
(unbiased) dark matter halos (10$^9$M$_\odot$ at $z=3$) to 
remove them from observed samples.

\acknowledgements %
We thank the organizers for running an enjoyable and educational workshop
and the editors for their hard work assembling this volume.  
This material is based upon work supported by the National Science 
Foundation under Grant. No. AST-0201667, 
an NSF Astronomy and Astrophysics Postdoctoral 
Fellowship (AAPF) awarded to E.G.

\end{document}